\documentclass[preprint,superscriptaddress]{revtex4-1}
\usepackage{graphicx}
\usepackage{amsmath}
\usepackage{amssymb}
\usepackage[colorlinks,linkcolor=blue,
            citecolor=blue,
            hyperindex,
            pdfstartview=FitH,
            plainpages=false]
            {hyperref}

\newcommand {\bisco}{Bi$_2$Sr$_2$CaCu$_2$O$_{8+\delta}$}
\newcommand {\uJcm}{$\mu$Jcm$^{-2}$}

\begin{document}
\title{Stimulated emission of Cooper pairs in a high-temperature cuprate superconductor}

\author{Wentao Zhang}
\email{wentaozhang@sjtu.edu.cn}
\affiliation{Materials Sciences Division, Lawrence Berkeley National Laboratory, Berkeley, California 94720, USA}
\affiliation{Department of Physics and Astronomy, Shanghai Jiao Tong University, Shanghai 200240, China}
\author{Tristan Miller}
\affiliation{Department of Physics, University of California, Berkeley, California 94720, USA}
\affiliation{Materials Sciences Division, Lawrence Berkeley National Laboratory, Berkeley, California 94720, USA}
\author{Christopher L. Smallwood}
\affiliation{Department of Physics, University of California, Berkeley, California 94720, USA}
\affiliation{Materials Sciences Division, Lawrence Berkeley National Laboratory, Berkeley, California 94720, USA}
\author{Yoshiyuki Yoshida}
\author{Hiroshi Eisaki}
\affiliation{Electronics and Photonics Research Institute, National Institute of Advanced Industrial Science and Technology, Ibaraki 305-8568, Japan}
\author{R. A. Kaindl}
\affiliation{Materials Sciences Division, Lawrence Berkeley National Laboratory, Berkeley, California 94720, USA}
\author{Dung-Hai Lee}
\author{Alessandra Lanzara}
\email{alanzara@lbl.gov}
\affiliation{Department of Physics, University of California, Berkeley, California 94720, USA}
\affiliation{Materials Sciences Division, Lawrence Berkeley National Laboratory, Berkeley, California 94720, USA}
\date {\today}

\begin{abstract}
The concept of stimulated emission of bosons has played an important role in modern science and technology, and constitutes the working principle for lasers. In a stimulated emission process, an incoming photon enhances the probability that an excited atomic state will transition to a lower energy state and generate a second photon of the same energy. It is expected, but not experimentally shown, that stimulated emission contributes significantly to the zero resistance current in a superconductor by enhancing the probability that scattered Cooper pairs will return to the macroscopically occupied condensate instead of entering any other state. 
Here, we use time- and angle-resolved photoemission spectroscopy to study the initial rise of the non-equilibrium quasiparticle population in a  Bi$_2$Sr$_2$CaCu$_2$O$_{8+\delta}$ cuprate superconductor induced by an ultrashort laser pulse. Our finding reveals significantly slower buildup of quasiparticles in the superconducting state than in the normal state. The slower buildup only occurs when the pump pulse is too weak to deplete the superconducting condensate, and for cuts inside the Fermi arc region. We propose this is a manifestation of stimulated recombination of broken Cooper pairs, and signals an important momentum space dichotomy in the formation of Cooper pairs inside and outside the Fermi arc region.

\end{abstract}

\maketitle

Numerous studies have utilized ultrafast infrared pulses to break Cooper pairs and create non-equilibrium quasiparticles (NEQPs) in conventional and high-temperature superconductors\cite{Gedik2004,Kaindl2005,Coslovich2011a,Smallwood2012,Zhang2013}. So far, most pump-probe spectroscopic studies on superconductors have focused on the recombination dynamics of NEQPs on picosecond time scale\cite{Gedik2005,Perfetti2007,Graf2011,Cortes2011,Smallwood2012}. Such phenomena is usually understood within the phenomenological Rothwarf-Taylor (RT) model, whose basic ingredient is simple bimolecular recombination\cite{Rothwarf1967,Gedik2004,Smallwood2012}.
On the contrary, processes like stimulated emission play a greater role in the faster femtosecond buildup dynamic and become negligent at picosecond time scale.
Important clues into what governs the short time scale dynamic may be provided by the emerging technique of time and angle resolved photoemission spectroscopy (trARPES), which adds simultaneous measurement of the superconducting gap recovery as well as momentum-dependent information about NEQP dynamics\cite{Smallwood2012a,Smallwood2014}.

Here we report a systematic trARPES study of the temperature, momentum, doping, and density dependence of the initial buildup of NEQPs in high temperature superconductor \bisco\ (Bi2212).  
In the superconducting state, and at low pump fluence, it takes $\sim 0.9$ ps for the NEQP population to reach the maximum.
In contrast, in the normal state, or when the pump fluence is sufficiently strong to deplete the entire superconducting condensate, the maximum in the NEQP population is reached immediately (within our time resolution 300 fs) after the arrival of the pump pulse. 
The NEQP population residing outside the Fermi arc also builds up almost instantly. Along the Fermi arc, the NEQP buildup time shortens as the energy gap increases.
We propose that these findings provide evidence of stimulated recombination of quasiparticles into the superconducting condensate.

\section*{Results}
\begin{figure}\centering\includegraphics{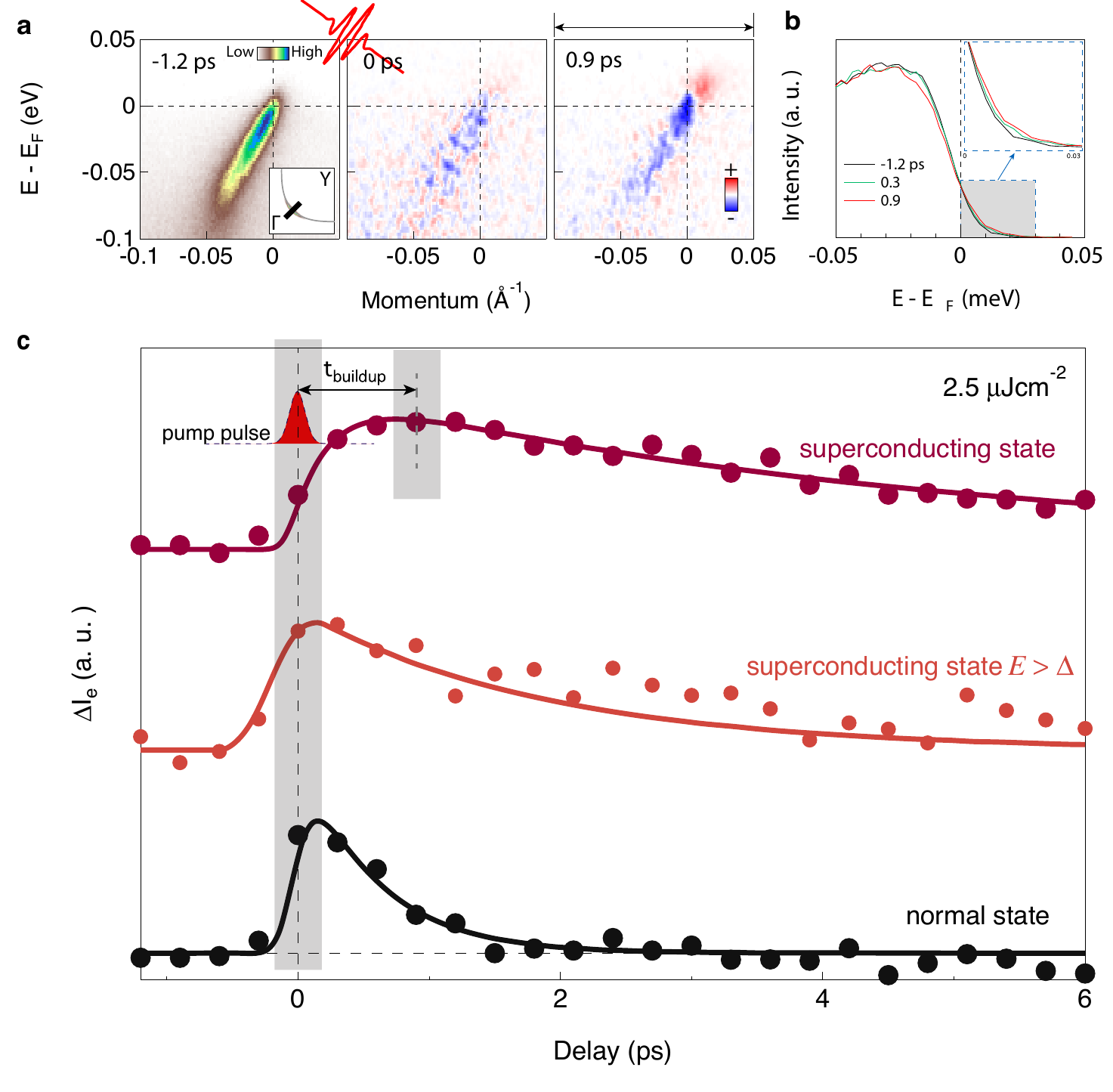}
\caption{
Buildup time in the nodal non-equilibrium electron excitations. Measurements were taken along a nodal cut ($\Gamma$(0, 0) -- Y($\pi$, $\pi$)) in an underdoped Bi2212 ($T_\text{c}=$ 78 K).
\textbf{a}, Equilibrium photoemission intensity at delay time $t=-1.2$ ps as a function of energy and momentum and pump-induced change of photoemission spectra as a function of energy and momentum at delay time 0 and 0.9 ps. The intensities as a function of energy by integrating the original photoemission spectra at delay time -1.2 ps, 0 ps and 0.9 ps are shown in the left panel. The measurement was taken at 20 K with a pump fluence of 2.4 \uJcm.
\textbf{b}, The integrated photoemission intensity across the arrowed momentum region for the three different delay time shown in (a).
\textbf{c}, The spectral gain $\Delta{I}_\text{e}$, by integrating the change of photoemission intensity above the Fermi level at 20 K, above the energy of maximum superconducting gap at 20 K, and above $T_\text{c}$ at 90 K as a function of delay time at pump fluence 2.4 \uJcm.
The width of the pump pulse in the cartoon is determined by frequency-resolved optical gating (FROG).
}
\label{Fig1}
\end{figure}

\textbf{Buildup of non-equilibrium states.}
Figure~\ref{Fig1} shows the buildup of the nodal NEQP population for an underdoped Bi2212 superconductor ($T_\text{c}$ = 78 K) above and below the superconducting transition temperature.
Fig.\ref{Fig1}a shows the equilibrium photoemission intensity map (before the arrival of the pump pulse) in the superconducting state and the pump-induced change of the same map for different delay times ($t=0$ represents the arrival time of the pulse).
Here the pump fluence (2.4 \uJcm) is far below the critical fluence ($\approx13$ \uJcm) needed to fully deplete the superconducting condensate\cite{Zhang2013}.  We observe a clear decrease in the photoemission intensity below the Fermi level, and an increase above the Fermi level, signifying the excitation of NEQPs.
However, the change in intensity is greater at $t=0.9$ ps than at $t=0$, indicating that the NEQP population requires a finite time to build up, and it can be evidenced in the integrated intensity as a function of energy shown in the left panel of Fig.\ref{Fig1}b.

Fig.\ref{Fig1}c shows the integrated change in the intensity above the Fermi level, defined as \cite{Perfetti2007,Cortes2011,Graf2011,Smallwood2012},
\begin{equation}
\label{IDeltaE}
\Delta{I}_\text{e}(t)=\int_0^Ed\omega\int_{-0.1}^{0.1}dk[I(\omega,k,t)-I(\omega,k,-1.2~\text{ps})]
\end{equation}
as a function of time  for T = 20 K $<$ T$_c$ and T = 90 K $>$ T$_c$.
Consistent with Fig. \ref{Fig1}a, the maximum change in intensity occurs at a much later time than $t=0$.  We will define the time at which $\Delta{I}_\text{e}$ is near maximum, $t_\text{buildup}$ as the ``build-up" time.
Later, we will describe a procedure to extract $t_\text{buildup}$.
For $t>t_\text{buildup}$ the decay rate of NEQP population above $T_\text{c}$ is much faster than that at below $T_c$, consistent with other reports in the literature\cite{Gedik2004,Gedik2005,Smallwood2012}.
From now on, we focus our attention on $\Delta{I}_\text{e}(t)$ for the usually overlooked $t\le t_\text{buildup}$. Strikingly, in the superconducting state, the maximum intensity change ($t_\text{buildup}$) is reached at $\sim$1 ps, while in the non-superconducting state the maximum intensity change is reached instantly.
If $\Delta{I}_e$ is integrated only above the maximum superconducting gap ($\Delta_\text{max}$), no delay in $t_\text{buildup}$ is observed.
In supplementary Figure 1 we demonstrate that the finite $t_\text{buildup}$ cannot be due to the limitation of our time resolution.

\begin{figure}\centering\includegraphics{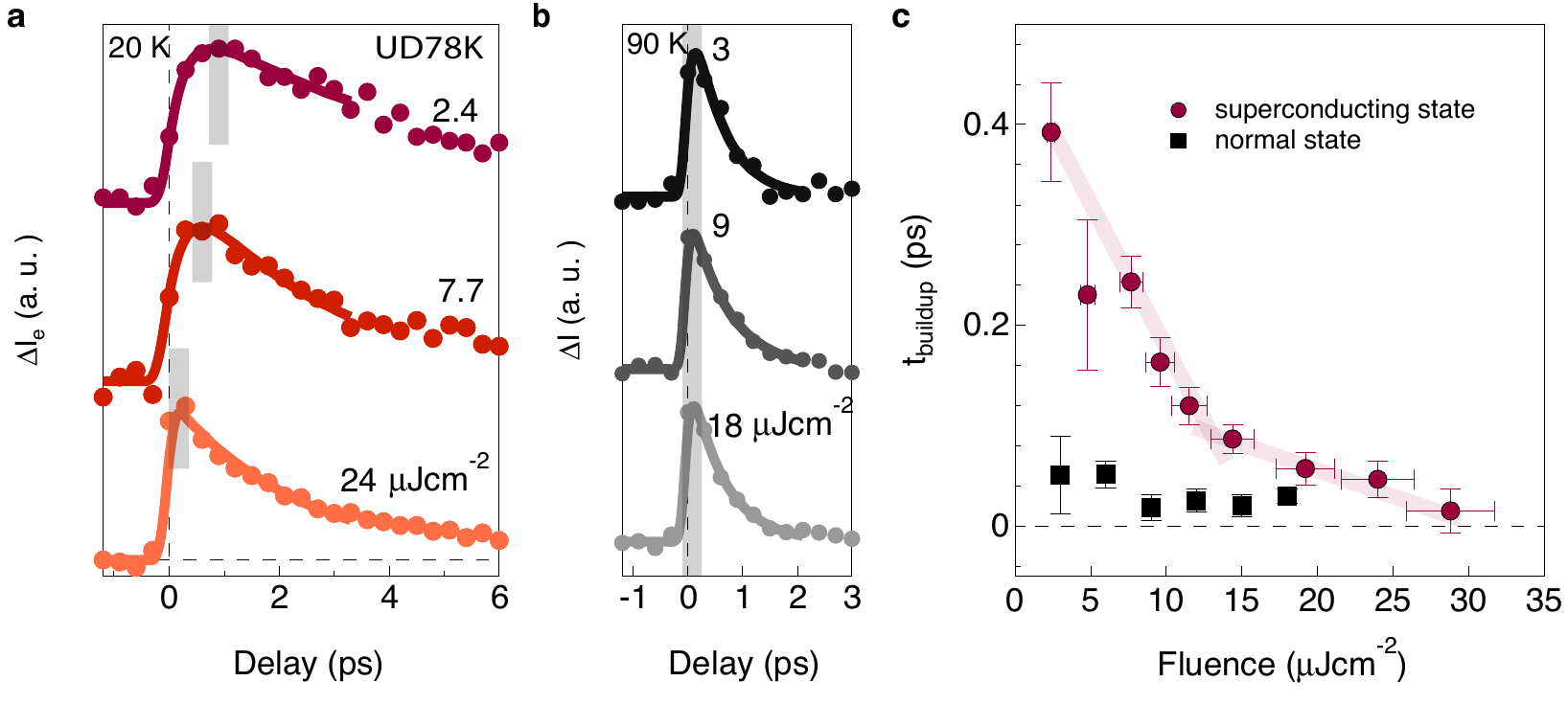}
\caption{
Pump fluence dependence of buildup time along nodal direction.
\textbf{a} and \text{b}, $\Delta I_\text{e}$ as a function of delay time and pump fluences at 20 K and 90 K, respectively, for an underdoped sample ($T_\text{c}=78$ K). The shaded line is a guide to the eyes.
\textbf{c}, The extracted buildup time $t_\text{buildup}$ from \textbf{a} as a function of pump fluence.
}
\label{Fig2}
\end{figure}

\textbf{Pump intensity dependence of the buildup of non-equilibrium states.}
Figure \ref{Fig2}a shows the low temperature ($T<T_c$)  $\Delta{I}_\text{e}(t)$ for different pump fluences. Very different buildup behaviors are observed for fluences below and above the critical fluence ($\approx13$ \uJcm). Specifically, 
at pump fluence equal to 24 \uJcm, $\Delta{I}_e(t)$ reaches its maximum instantly within our time resolution. This fast buildup is similar to the normal state data in Fig. \ref{Fig1}c. On the other hand at lower pump fluence, $t_\text{buildup}$ appears as large as 0.9 ps.
In marked contrast with the low temperature data, for $T>T_c$ the buildup of non-equilibrium electrons does not show a significant delay for any pump fluence studied (Fig. \ref{Fig2}b). 

To precisely define the buildup time, $t_\text{buildup}$, we fit $\Delta{I}_\text{e}(t)$  using a resolution-convolved function given by
\begin{equation}
\label{DecayOnsetFunction}
\Delta{I_\text{e}}(t)=A({1-e^{-t/t_\text{buildup}}})e^{-\gamma{t}}\otimes Gauss(t, \text{FWHM})
\end{equation}
where $\gamma$ is the long-time NEQP population decay rate.  The Gaussian accounts for the experimental time resolution, and its full width at half maximum (FWHM) is 300 fs. Equation \ref{DecayOnsetFunction} captures the dynamics of the initial buildup of non-equilibrium quasiparticles quite well (see solid lines in Fig. \ref{Fig2}a). Fig. \ref{Fig2}c shows the extracted $t_\text{buildup}$ as a function of pump fluence.  At low temperature, $t_\text{buildup}$ is $\sim$400 fs at the lowest fluence, and decreases as the fluence increases until it saturates at the normal state value of $<$ 100 fs.  The saturation occurs at the critical fluence required to deplete the superconducting condensate\cite{Zhang2013}, but note that the superconducting gap does not close until about 0.6 ps later\cite{Smallwood2014}.
The absence of a buildup time above $T_\text{c}$ or above the critical fluence (Fig. \ref{Fig2}a) suggests a connection between the slow buildup and the presence of the superconducting condensate.

\begin{figure}\centering\includegraphics{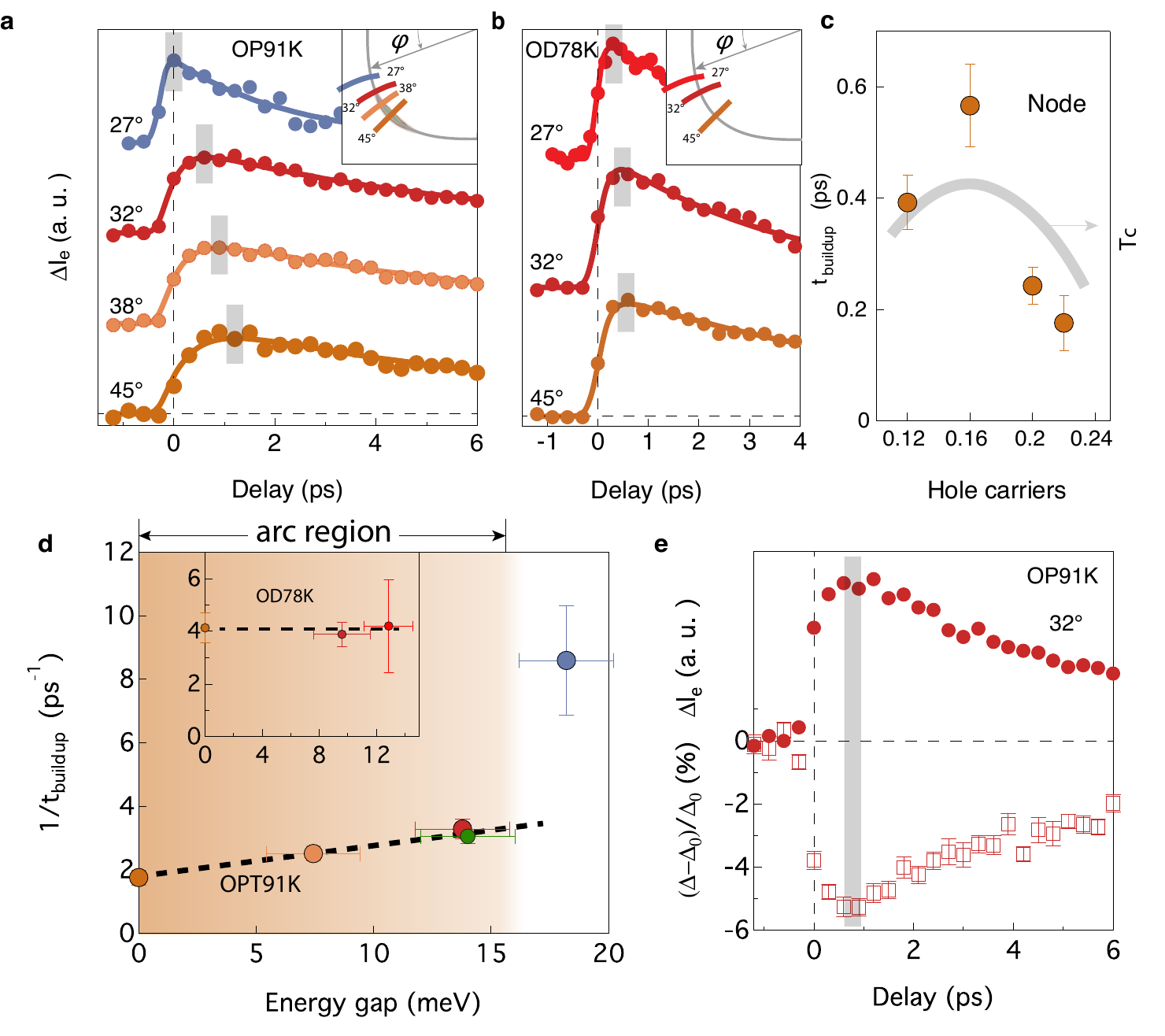}
\caption
{
Momentum dependence of the buildup time. The measurements were taken at 20 K on nearly optimally ($T_\text{c}=$91 K)  and overdoped  ($T_\text{c}=$78 K)  Bi2212 samples with pump fluence of 2.4 \uJcm.
\textbf{a} and \textbf{b}, $\Delta I_\text{e}$ as a function of delay time measured inside (red curves) and outside (blue curve) Fermi arc for optimally doped sample (panel a) and inside the Fermi arc for an overdoped samples (panel b).
\textbf{c}, Doping dependence of the buildup time at the node.
\textbf{d}, The extracted buildup rate $1/t_\text{buildup}$ from \textbf{a} as a function of energy gap at the Fermi momentum for four cuts shown in inset of \textbf{b}. The green data point is a separate measurement at the Fermi surface angle of $32^\circ$. The brown region
\textbf{e}, Comparison between $\Delta I_\text{e}$ and the change of energy gap as a function of delay time at the Fermi surface angle of $32^\circ$.
}
\label{Fig3}
\end{figure}

\textbf{Momentum dependence of the buildup of non-equilibrium states.}
In momentum space, similar delayed buildup is observed in $\Delta I_e(t)$ throughout the Fermi arc region.  Fig. \ref{Fig3}a shows the momentum-dependent buildup time in an optimally doped sample in response to a weak (2.4 \uJcm) pump.  As previously reported\cite{Smallwood2012}, the long-time decay rate ($\gamma$) of the NEQP population increases from node to off-node.  Inside the Fermi arc region, there is a delay in buildup, which decreases away from the node. In contrast, for a momentum cut outside the Fermi arc region, there is no delay, and $t_\text{buildup}$ drops to the normal state value of about 100 fs.
We have also compared the buildup time for the NEQP population in samples with different doping level (overdoped sample, Fig. \ref{Fig3}b), and the same momentum dependence of $t_\text{buildup}$ is observed: the build up time is non zero only on the Fermi arc.
For the overdoped sample, all the momentum cuts go through the Fermi arc region and hence they all show an apparent delay in the buildup of the NEQP population.
However, compared with the optimally- and under-doped samples, the delay is less pronounced (Fig. \ref{Fig3}c).
The momentum dependence of the build up rate for the NEQP population is shown in Fig. \ref{Fig3}d.
For each momentum we extract the corresponding gap value. The NEQP is then plotted as a function of gap size.
Gap values are extracted using a standard procedure of fitting the symmetrized energy distribution curves at the Fermi momentum to phenomenological single particle spectral functions\cite{Norman1998} (supplementary Figure 2 and supplementary Discussion 2). 
It is clear that there is a distinct mechanism for the buildup time for the states beyond the Fermi arc region.

Note that the difference cannot be attributed to the lack of a gap inside the Fermi arc region, since the Fermi arc itself does not develop until about 0.6 ps after pumping\cite{Smallwood2014}.
Interestingly, the build up time of the NEQP population is identical to that of the photoinduced shift of the superconducting gap at the same momentum (Fig. \ref{Fig3}e). The similar timescale further suggests that the superconducting condensate is a key ingredient in establishing the build up time.

\section*{Discussion}
In the following we propose an explanation of these observations that takes into account the quantum coherence of the Cooper
pair condensate. When the pump pulse photoexcites the material, the input energy is partially stored in the electronic degrees of freedom and partially in the lattice degrees of freedom. Where there is a superconducting condensate, those NEQPs whose time reversal partner is also excited can quickly recombine into Cooper pairs due to a process of coherent recombination stimulated by the existing pair condensate. This process serves to reduce the proportion of input energy initially stored in electronic degrees of freedom, hence causing the observed delay in the buildup of the NEQP population. Such a stimulated recombination process is absent when the superconducting condensate is depleted by either the temperature or the high pump fluence. After $t_\text{buildup}$, what are left are ``unpaired'' excited quasiparticles, meaning the time reversed partners of these quasiparticles are not simultaneously excited. At a longer time scale, due to momentum-changing impurity, some subset of those ``unpaired'' excited quasiparticles will change into ``paired'' ones and quickly recombine. The bottleneck of the recombination process is the conversion of unpaired NEQPs into paired ones, which is considerably slower\cite{Rothwarf1967,Gedik2004,Smallwood2012}. This process is depicted as real space diffusion triggered ``bimolecular'' recombination.

\begin{figure}\centering\includegraphics{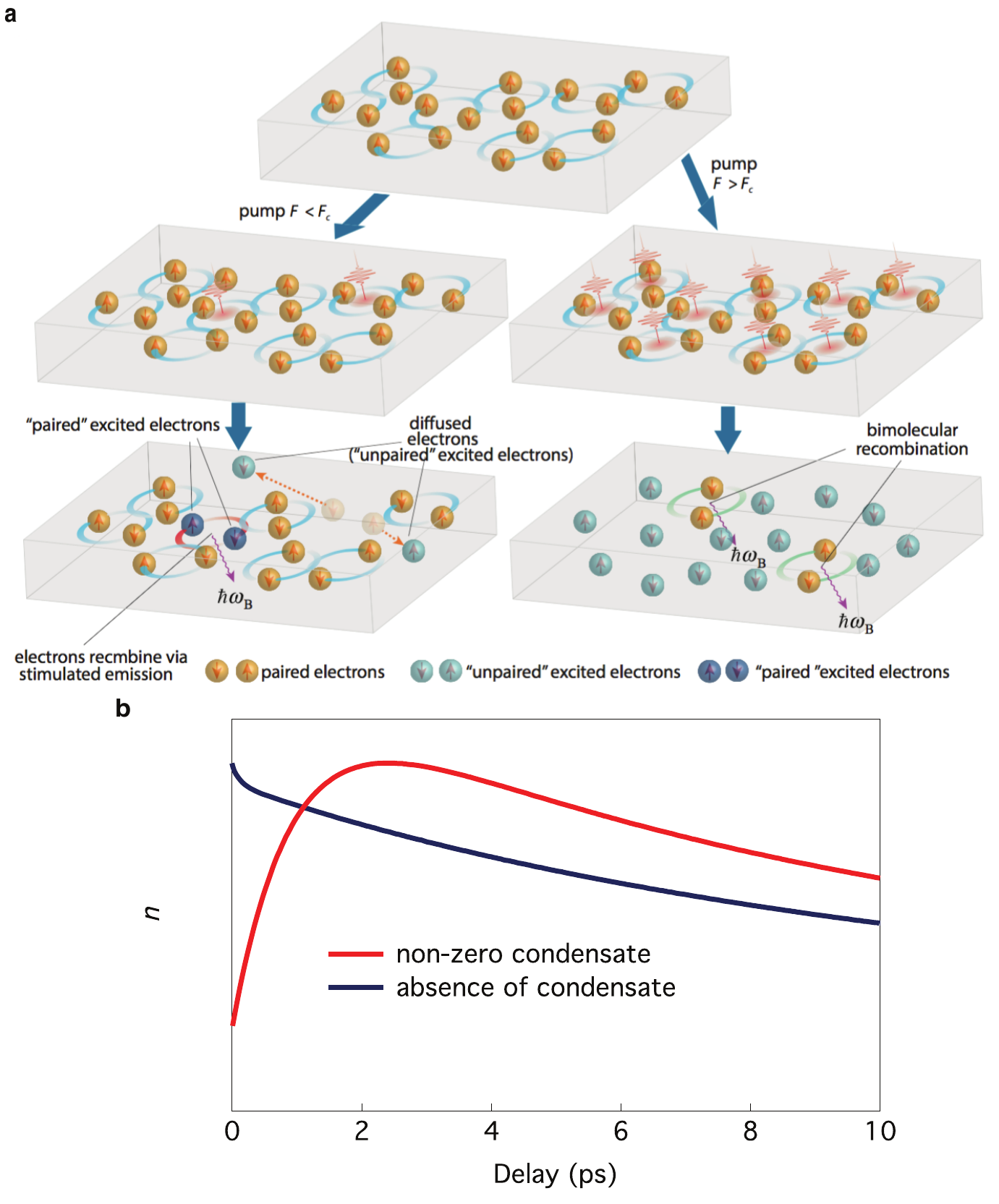}
\caption{
Response of a superconductor to pump photons.
\textbf{a}, Cartoons to show the response of a superconductor to a pump laser for fluences lower and higher than the critical fluence.
\textbf{b}, Simulations of the numbers of non-equilibrium electrons ($n$) as a function of delay time. A non-zero buildup time is observed only in the presence of a condensate.
The curves are the numerical solutions of the equations in supplementary Discussion 3 using the same parameters as shown in supplementary Figure 3b but with $\gamma_\text{p}=0$ for the absence of the stimulated emission of Cooper pairs into the condensate and $\gamma_\text{p}=1000$ for non-vanishing stimulated emission.
}
\label{Fig4}
\end{figure}

A cartoon of of this physical scenario is shown in Fig. \ref{Fig4}a.  Note that this explanation is similar to that of earlier optical spectroscopy studies, in that the finite buildup time arises because much of the initial photoexcitation energy is stored in the lattice degrees of freedom\cite{Demsar2003,Kusar2008,Beck2011}.
However, the main difference is that we attribute this initial energy specifically to the presence of the superconducting condensate.  This is a key aspect to account for the temperature, energy and fluence dependence here reported, which otherwise cannot be explained.

The importance of including stimulated recombination into the RT equations is summarized in Fig. \ref{Fig4}b, where the delay dependence of the number of non-equilibrium electrons is shown.  Without superconducting condensate (black curve) no buildup time is observed.  However, when the RT equations are augmented to include the stimulated recombination channel (see Supplementary Figure 3 and Discussion 3), the results are qualitatively consistent with the experimental observations (see red curve in Fig. \ref{Fig4}b) . Note that due to the electronic inhomogeneity of Bi-2212 the momentum conservation rules are relaxed, i.e. when ``paired'' NEQP recombine they can emit phonons (bosons) with a sufficiently wide range momentum.
Within this scenario, the momentum dependence of the buildup time can be easily accounted for by the momentum dependence of the recombination rate ($\gamma$), that increases as we move away from the node\cite{Smallwood2012}. 
Additionally, the absence of a buildup time beyond the Fermi arc tip, put a strong constraint on the presence of a condensate in this momentum region, another important manifestation of the nodal off nodal dichotomy\cite{Fu2006}.

One alternate interpretation of our results is that initial NEQPs far from the node scatter towards the node, allowing the NEQP population near the node to build up over a longer period of time.  Such scattering would be induced by the anisotropic form of the superconducting gap.  However, this interpretation does not account for the absence of a buildup time at high fluences or high temperatures.  Above the critical fluence, the superconducting gap  is not suppressed until about 0.6 ps after the pump pulse\cite{Smallwood2014}, some time after the NEQP population has reached its maximum, and above $T_c$, there is still an anisotropic pseudogap. 

To conclude, our main finding is that upon photoexcitation the delay in the buildup of non-equilibrium quasiparticle population in higher temperature superconductors is sensitive to whether there is a superconducting condensate. We propose that our observation reveals the stimulated recombination of photoexcited quasiparticles.

\section*{Methods}
Measurements were taken on our home-built trARPES system.
The experimental setup is the same as that reported previously\cite{Graf2011,Smallwood2012,Smallwood2012a,Zhang2013}.
Transient states in the solids are created by pumping with infrared laser pulses ($h\nu=1.48$ eV), then they are probed by a following ultra-violet pulse ($h\nu=5.93$ eV) using ARPES setup equipped with a SPECS PHOIBOS 150 electron spectrometer.
Time resolution is achieved by varying the delay time between the probe and pump pulses, with a total resolution of $\sim$300 fs determined by a cross-correlation of pump and probe pulses measured on hot electrons.
The total energy resolution is 23 meV characterized by fitting the Fermi edge to a resolution-convolved Fermi-Dirac distribution function.
Samples with superconducting transition of underdoped ($T_\text{c}=78$ K, UD78K) and optimally doped ($T_\text{c}=91$ K, OP91K) and overdoped ($T_\text{c}=78$ K, OD78K) were studied.
The samples were cleaved $\emph{in situ}$ at 20 K in ultra high vacuum with a base pressure lower than $5\times10^{-11}$ mbar.

\begin{acknowledgments}
This work was supported by Berkeley Lab's program on Ultrafast Materials Sciences, funded by the U.S. Department of Energy, Office of Science, Office of Basic Energy Sciences, Materials Sciences and Engineering Division, under Contract No. DE-AC02-05CH11231.
\end{acknowledgments}

\section*{Author contributions}
W. Z. conceived the project and the experiments together with A. L. The experiments were taken by W. Z., T. M. and C. L. S. Samples were grown and characterized by Y. Y. and H. E. The manuscript was written by W. Z., A. L. and D. H. L. with input from all authors.

\section*{Additional information}
The authors declare no competing financial interests.

\newpage

\newpage

\newpage

\newpage

\end{document}